\documentclass[runningheads]{llncs}
\usepackage{amssymb}
\usepackage{graphicx}
\usepackage{url}

\begin{document}

\mainmatter  

\def\v#1{\mathbf{#1}}
\def\arccosh{\mathop\mathrm{arccosh}}
\def\dotproduct#1#2{{\langle {#1}, {#2} \rangle}}
\def\ceil#1{ {\lceil{#1}\rceil} }
\def\ds{\mathrm{d}s}
\def\dt{\mathrm{d}t}
\def\dx{\mathrm{d}x}
\def\dy{\mathrm{d}y}
\def\arctanh{\mathop\mathrm{arctanh}}
\def\floor#1{ {\lfloor {#1} \rfloor} }
\def\ballE{B}
\def\dotproduct#1#2{{\langle {#1}, {#2} \rangle}}
\def\vor{\mathrm{vor}}
\def\calX{\mathcal{X}}
\def\calP{\mathcal{P}}

\title{Hyperbolic Voronoi diagrams made easy}

\author{Frank Nielsen\inst{1} \and Richard Nock\inst{2}}
\institute{\'Ecole Polytechnique / Sony Computer Science Laboratories Inc\\
Palaiseau, France / Tokyo, Japan\\
{\tt nielsen@lix.polytechnique.fr}
\and CEREGMIA-UAG\\
University of Antilles-Guyane\\
Martinique, France\\
{\tt rnock@martinique.univ-ag.fr}
}
\maketitle

\begin{abstract}
We present a simple framework to compute   hyperbolic Voronoi diagrams of   finite point sets as affine diagrams.
We prove that bisectors in Klein's non-conformal disk model are hyperplanes that can be interpreted as power bisectors of Euclidean balls.
Therefore our method simply consists  in computing  an  equivalent clipped power diagram followed by a mapping transformation depending on the selected representation of the hyperbolic space (e.g., Poincar\'e conformal disk or upper-plane representations). 
We  discuss on extensions of this approach to weighted and $k$-order diagrams, and describe their dual triangulations.
Finally, we consider two useful primitives on the hyperbolic Voronoi diagrams for designing tailored user interfaces of an image catalog browsing application in the hyperbolic disk:
(1) finding   nearest neighbors, and (2) computing  smallest enclosing balls.
\end{abstract}

\section{Introduction and prior work}
The birth and spread of science was originally initiated by geometry, the science of (Earth) measurements.
Euclid (300 BC) laid the foundations of geometry in his masterpiece book series {\it Elements}
 that still remain the basis of mathematics nowadays, two millenia later.
  Euclidean geometry was assumed to be the only consistent geometry until the nineteenth century.
 Eventually,  failing to prove Euclid's fifth postulate  (known as the parallel postulate) from the others opened the door to abstract geometries.
 The  first two abstract geometries that were historically called imaginary geometries are the hyperbolic geometry and the spherical   geometry.
 The essential difference between Euclidean and these non-Euclidean geometries is the nature of parallel lines. 
 While in Euclidean geometry there is a unique line passing through a given point and parallel to another line, there can be infinitely many in  the hyperbolic geometry~\cite{beltrami-1868} and none in spherical/elliptical geometries. 
In this paper, we focus on characterizing and building efficiently the Voronoi diagram in hyperbolic geometry. 
 
The Voronoi diagram~\cite{compgeom-1998} of a finite point set partitions the space in cells according to proximity relations induced by a distance function $d$.
The structures of Voronoi diagrams appear in various fields such as crystallography, astronomy and biology.
Let $\calP=\{\v{p}_1, ..., \v{p}_n\}$ be a set of $n$ $d$-dimensional vector points of space $\calX$. Then the Voronoi diagram of $\calP$ is defined by the collection of proximal regions $\vor(\v{p}_i)$'s, called Voronoi cells,   such that
$
\vor(\v{p}_i)=\{ \v{x} \in \calX\ |\  d(\v{p}_i,\v{x}) \leq d(\v{p}_j,\v{x}),\ \forall \v{p}_j\in\calP \}
$.

The Voronoi diagram in hyperbolic geometry has already been partially studied;
Onishi and Takayama~\cite{voronoidiagram-upperplane-1996} investigated the hyperbolic Voronoi diagram construction in the Poincar\'e upper half-plane.
Onishi and Itoh~\cite{voronoidiagram-hadamard-2002} further extended the Voronoi diagram in Hadamard manifold that are simply connected complete manifold with non-positive curvature. 
Boissonnat et al.~\cite{DBLP:journals/ijcga/BoissonnatCDT96}  considered  the hyperbolic Voronoi diagram in the Poincar\'e conformal upper plane and the Poincar\'e conformal disk.\footnote{See Figure~6 and Figure~8 of~\cite{DBLP:journals/ijcga/BoissonnatCDT96} for illustrations.}
Boissonnat and Yvinec  described in their computational geometry textbook the hyperbolic Voronoi diagram in the Poincar\'e $d$-dimensional half-space $\mathbb{H}^d_+$~\cite{compgeom-1998}, pp. 449-454.  They do not introduce  explicitly the hyperbolic distance but rather show how to answer proximity queries (e.g., whether $B$ or $C$ is closer to $A$) using the notion of pencil of spheres. They deduced that the complexity of the hyperbolic Voronoi diagram is $O(n\log n+n^{\ceil{\frac{d}{2}}})$.
Nilforoushan and Mohades~\cite{DBLP:conf/iccsa/NilforoushanM06} concentrated on the hyperbolic Voronoi diagram in the Poincar\'e 2D disk for which they  report an incremental quadratic algorithm, and characterize geometrically the orthogonality of bisectors with geodesics.
Note that by using the graphics processor unit (GPU), it is easy to rasterize interactively Voronoi diagrams~\cite{n-voronoigpu-2008} for any arbitrary distance function, including the distance of hyperbolic geometry. 

In this paper, we revisit these work by considering the Klein projective disk model.
We showed in section~\ref{sec:hvdpd} that bisectors are hyperplanes, implying that the Voronoi diagram is affine and therefore can be easily 
constructed from an equivalent power diagram. We report the one-to-one conversion formula to change the representation of the hyperbolic geometry: Klein or Poincar\'e disk models and Poincar\'e upper half plane model. We further characterize the weighted and $k$-order diagrams in the Klein disk model and explain the dual hyperbolic Delaunay triangulation.
In section~\ref{sec:app}, we present an image browsing application in the hyperbolic disk that requires two basic user interface selection operations that are solved by means of nearest neighbor queries and by finding the smallest enclosing ball in hyperbolic geometry.

\section{Hyperbolic Voronoi diagrams from power diagrams\label{sec:hvdpd}}

\subsection{Klein's non-conformal disk model}
The Klein model (or Beltrami-Klein model~\cite{beltrami-1868}) uses the interior of a unit circle for fully representing the hyperbolic plane.
The Klein model is also known as the projective disk model where lines are depicted by chords of the circle (e.g., line segments joining any two points of the circle).
Although the Klein model offers a simple visualization of geodesics as line segments, it has the disadvantage of being a non-conformal representation:
That is, angles of the hyperbolic plane are not preserved, and get distorted in this model.
(That explains why the Poincar\'e conformal disk representation is often preferred.)
The distance between any two points $\v{p}$ and $\v{q}$  of the Klein disk with Euclidean coordinates ($||\v{p}|| < 1$ and $||\v{q}|| < 1$) is computed as

\begin{equation}\label{eq:dist}
h_{\mathbb{K}}(\v{p},\v{q})=\arccosh \frac{1-\dotproduct{\v{p}}{\v{q}}}{\sqrt{(1-||\v{p}||^2)(1-||\v{q}||^2)}},
\end{equation}
where $\arccosh x=\log (x+\sqrt{x^2-1})$.

\subsection{Bisectors in Klein projective model are hyperplanes}
Bisectors are core primitives to characterize and compute Voronoi diagrams.
The bisector $B$ of two points $\v{p}$ and $\v{q}$ with respect to a given distance function $D$ is defined as the loci of the points $\v{c}$ that are at the same distance of to $\v{p}$ and $\v{q}$: $B: \{ \v{c} \ |\ D(\v{p},\v{c})=D(\v{q},\v{c})\}$.
Let us derive from the distance function of Eq.~\ref{eq:dist} the  equation of the  bisector in Klein's disc model of hyperbolic geometry: $h_{\mathbb{K}}(\v{p},\v{c})=h_{\mathbb{K}}(\v{q},\v{c})$.
That is
\begin{eqnarray}
\frac{1-\dotproduct{\v{p}}{\v{c}}}{\sqrt{(1-||\v{p}||^2)(1-||\v{c}||^2)}}=\frac{1-\dotproduct{\v{c}}{\v{q}}}{\sqrt{(1-||\v{c}||^2)(1-||\v{q}||^2)}},  \nonumber \\
\dotproduct{\v{c}}{\sqrt{1-||\v{p}||^2}\v{q}-\sqrt{1-||\v{q}||^2}\v{p}} + \sqrt{1-||\v{q}||^2} - \sqrt{1-||\v{p}||^2}=0 \label{eq:kb}
\end{eqnarray}
\noindent This is a linear equation $\dotproduct{\v{a}}{\v{c}}+b=0$ in $\v{c}$, with   $\v{a}=\sqrt{1-||\v{p}||^2}\v{q}-\sqrt{1-||\v{q}||^2}\v{p}$ and $b=\sqrt{1-||\v{q}||^2}-\sqrt{1-||\v{p}||^2}$. Thus the bisector is a hyperplane, and the Voronoi diagram is said {\em affine}~\cite{powerdiagrams-1987}.

\subsection{Hyperbolic Voronoi diagrams as equivalent power diagrams}

The power distance of a point $\v{x}$ to a Euclidean ball
$B=\ballE(\v{p},r)$ is defined as
$||\v{p}-\v{x}||^2-r^2$. Given $n$ balls $B_i=\ballE
(\v{p}_i, r_i)$, $i=1,\ldots ,n$, the {\em power diagram} (also called Laguerre diagram) of the $B_i$'s is
defined as the {\it minimization diagram} of the corresponding $n$ functions
$D_i(\v{x} )=||\v{p}_i-\v{x}||^2-r^2$. 
The power bisector of
any two balls $\ballE(\v{p}_i,r_i)$ and $\ballE(\v{p}_j,r_j)$ is the {\em radical
hyperplane}~\cite{powerdiagrams-1987} of equation
\begin{equation}\label{eq:bipd}
2\dotproduct{\v{x}}{\v{p}_j-\v{p}_i}+||\v{p}_i||^2-||\v{p}_j||^2+r_j^2-r_i^2=0.
\end{equation}

Thus power diagrams are also affine diagrams.
Interestingly, Aurenhammer~\cite{powerdiagrams-1987} proved that {\it all} affine diagrams can be constructed from equivalent power diagrams.
That is power diagrams are universal representations of affine diagrams.
Therefore any algorithm to build power diagrams can also build all affine Voronoi diagrams once the equivalence relations are explicitly given.
Thus we need to report the mapping that associates to points in the Klein disk  corresponding Euclidean balls so that their bisectors match. 
By identifying the bisector equation Eq.~\ref{eq:kb} with that of Eq.~\ref{eq:bipd}, we find that  point $\v{p}_i$ is mapped to a ball of center 
$\v{c}_i=\frac{\v{p}_i}{2\sqrt{1-\|\v{p}_i\|^2}}$
 and radius 
$r_i=\frac{\|\v{p}_i\|^2}{4(1-\|\v{p}_i\|^2)}-\frac{1}{\sqrt{1-\|\v{p}_i\|^2}}$.
Note that some of the radii may be negative (imaginary numbers, e.g. $\|\v{p}\|^2=0\rightarrow r=-1=i^2$), i.e. represent imaginary balls.
A careful analysis shows that the radius is negative for $\|\v{p}\|^2<\sqrt{ 8(\sqrt{\frac{5}{4}-1)} }\simeq 0.9718...$ and positive otherwise.

\begin{theorem}
The hyperbolic Voronoi diagram of $n$ $d$-dimensional points can be obtained  in the Klein disk model as an equivalent power diagram.
The diagram has combinatorial complexity $O(n^{\ceil{\frac{d}{2}}})$ and can be built  in
$O(n\log n+n^\ceil{\frac{d}{2}})$ time. 
\end{theorem}

Since power diagrams are defined on the full space $\mathbb{R}^d$, the equivalence also allows us to define the hyperbolic Voronoi diagram on the boundary circle.
Figure~\ref{fig:hvdfrompd} depicts the construction of the hyperbolic Voronoi diagram of a point set as the corresponding power diagram of associated balls. Note that the equivalence is only one way as there exists power diagrams with empty cells, which is not possible for the hyperbolic Voronoi diagrams.

\def\ttt{5cm}
\begin{figure}
\centering
\begin{tabular}{cc}
Hyperbolic Voronoi (Klein disk) & Equivalent power diagram of balls\\ \hline
\includegraphics[bb=0 0 1024 1024, width=\ttt]{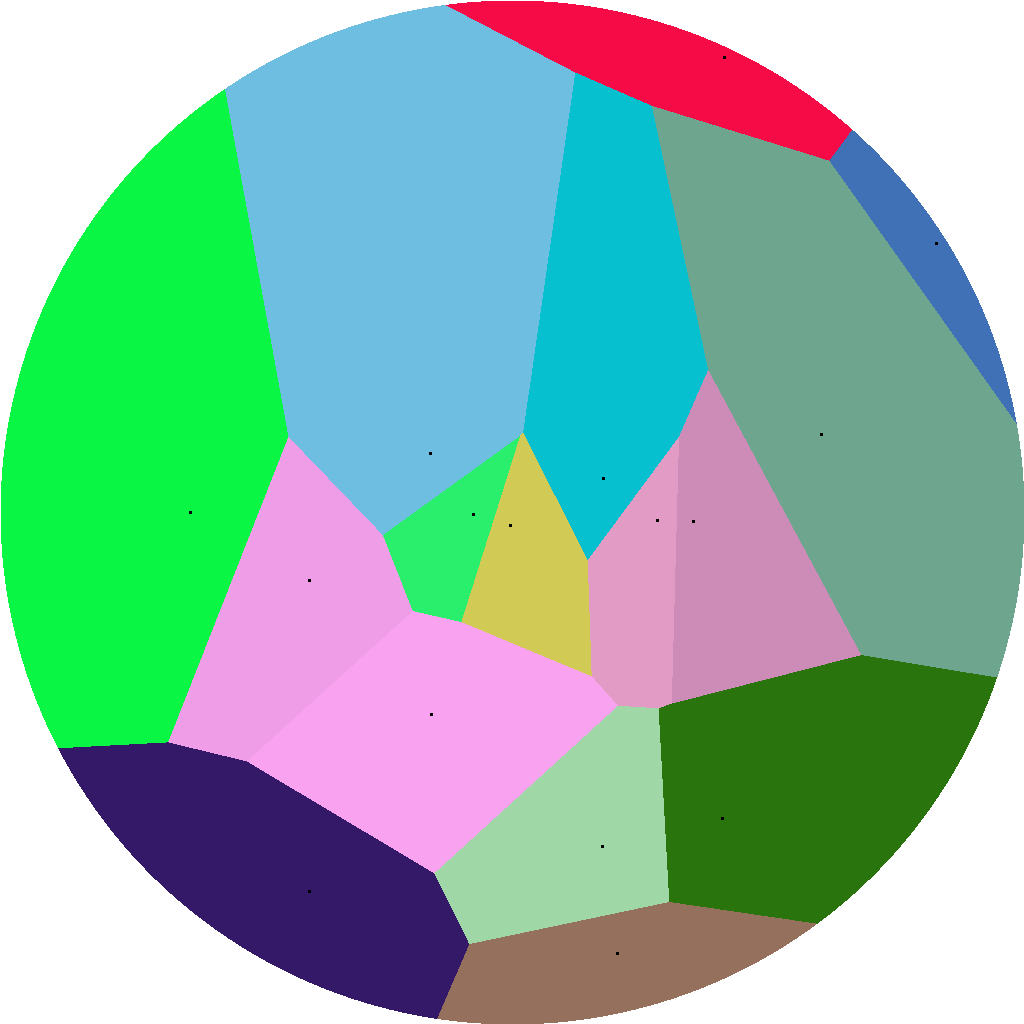} &
\includegraphics[bb=0 0 1024 1024, width=\ttt]{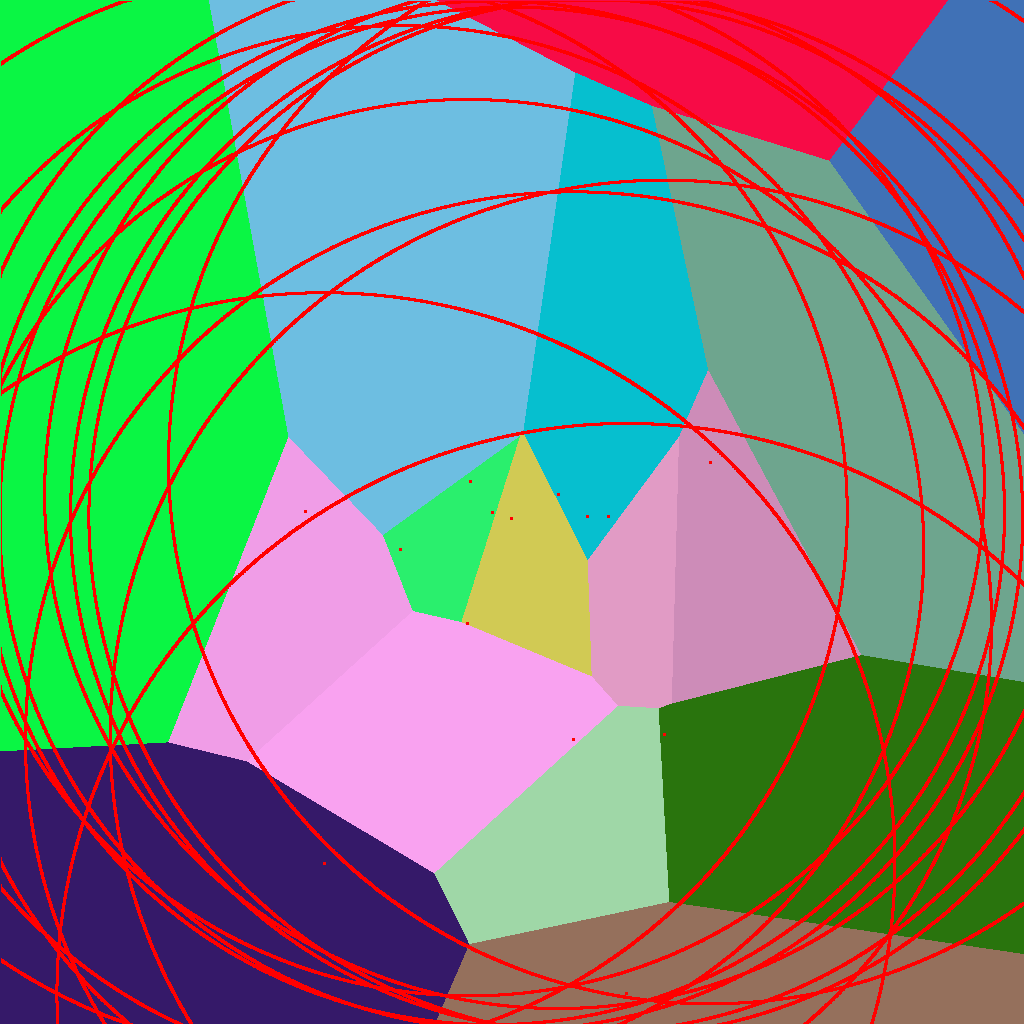}\\ \hline
\end{tabular}

\caption{(Left) Hyperbolic Voronoi diagram in the Klein projective disk model is affine, and can thus be built equivalently from (right) a clipped  power diagram of corresponding balls. Notice that the power diagram fully covers the Euclidean space.
\label{fig:hvdfrompd}
}
\end{figure}

Building efficiently power diagrams is a well-studied problem~\cite{powerdiagrams-1987} for which many highly optimized implementations are available.
For example, CGAL\footnote{The computational geometry algorithms library, see \url{http://www.cgal.org}} provides this module.

\subsection{Converting  Klein/Poincar\'e  disk models}
 
The hyperbolic Voronoi diagram in the  Poincar\'e disk $\mathbb{P}^2$ is conformal, i.e. it preserves the notion of hyperbolic angles defined by intersecting curves in the representation space (the embedding space called the model). 
 Straight lines in the hyperbolic geometry are represented by arcs of circles perpendicular to the unit circle in the Poincar\'e disk. 
 Two non-intersecting arcs denote parallel hyperbolic lines, and two arcs intersecting orthogonally correspond to perpendicular lines.
 There exists infinitely many geodesics passing through a given point parallel to a given geodesic, contradicting Euclid's fifth postulate of Euclidean geometry. 
The distance between any two vector points $\v{p}$ and $\v{q}$ with Euclidean coordinates is 
$h_\mathbb{P}(\v{p},\v{q})=\arccosh (1+\delta(\v{p},\v{q}))$ with 
$\delta(\v{p},\v{q})=2\frac{||\v{p}-\v{q}||^2}{(1-||\v{p}||^2)(1-||\v{q}||^2)}$.
 A geodesic (line segment in hyperbolic geometry) is expressed in the Poincar\'e disk by either a portion of a circle
 $x^2+y^2-2(ax+by)+1=0$ (with $a^2+b^2>1$) or by a line $ax=by$.

A point $\v{k}$ in the Klein disk corresponds to a point $\v{p}$ in the Poincar\'e disk (and vice-versa) using the following radial scaling formula
$\v{p}=\frac{1-\sqrt{1-\dotproduct{\v{k}}{\v{k}}}}{\dotproduct{\v{k}}{\v{k}}} \v{k}$ and $\v{k}=\frac{2}{1+\dotproduct{\v{p}}{\v{p}}}\v{p}$.

\begin{figure}
\centering
\includegraphics[bb=0 0 512 510, width=0.4\textwidth]{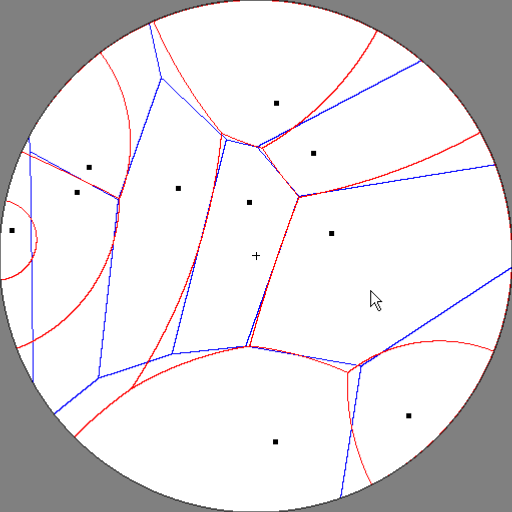}\hfill
\includegraphics[bb=0 0 509 507, width=0.4\textwidth]{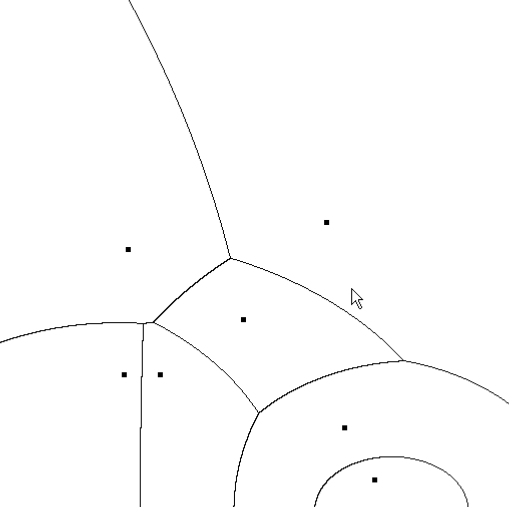}
\caption{\label{fig:KP} Hyperbolic Voronoi diagram in the disk and upper plane models: 
(left) Poincar\'e conformal Voronoi delimited by circular arcs superposed with the equivalent Klein affine Voronoi diagram, (b) Poincar\'e upper plane  conformal Voronoi diagram.}
\end{figure}

The relationship between the two Klein/Poincar\'e models is therefore a projection from the center of the disk.
Figure~\ref{fig:KP} shows both Klein affine and Poincar\'e circular arc Voronoi diagrams.
The straight line  passing through two points in the Klein model intersects the unit circle in two ideal points, with the
corresponding Poincar\'e line being a circular arc orthogonal to the boundary disk.

\subsection{Sphere inversions and M\"obius transformations}
M\"obius transformations (also called homographies) form a geometric group defined by inversions of spheres.
An inversion of the space by a sphere centered at $O$ with radius $r$, maps to itself all rays emanating from $C$ such that the product of a point on that ray with its image (i.e., the mapped point) equals $r^2$.
  M\"obius transformations  preserve angles as well.
The isometries of hyperbolic geometries are in one-to-one mapping with M\"obius transformations.
Thus we can navigate hyperbolic space by smoothly moving the viewpoint of the Poincar\'e disk model using M\"obius transformations (see Section~\ref{sec:ui}).

Note that in computational geometry, the M\"obius Voronoi diagram~\cite{moebius-2003} of a point set $\{\v{p}_i\}_i$ is defined as the space partition  relating proximity structure induced by distance functions $D_i(\v{x})=\lambda_i\|\v{x}-\v{p}_i\|^2-\mu_i$, where the $\lambda_i$'s are multiplicative factor constants and the $\mu_i$'s are additive constants (i.e., additively and multiplicatively weighted Voronoi diagrams).
 Hyperbolic Voronoi diagrams are special cases of M\"obius diagrams that can be viewed as affine diagrams. 

\subsection{Converting to Poincar\'e upper half plane}
To convert the affine Klein Voronoi diagram to the Poincar\'e upper half plane $\mathbb{H}_+^2$ model, we first convert the Klein disk model to the Poincar\'e disk model as explained in the previous section. 
We then interpret 2D points $(a,b)$ as complex numbers $z=a+ib$ in the complex space, so that the Poincar\'e disk model is the space $\{z\in\mathbb{C}\ |\ |z|<1\}$ and the upper half plane model corresponds to the space $\{z=a+ib\in\mathbb{C}\ |\ b>0\}$. Both Poincar\'e disk/upper plane models are conformal.
We then apply the following M\"obius transformation
$f: \mathbb{C}\cup\{\infty\} \rightarrow \mathbb{C}\cup\{\infty\}$ with $z\mapsto f(z)=\frac{i(z+1)}{1-z}$.
Reciprocally, to convert the upper half plane to the disk model, we apply the inverse M\"obius transformation 
$f^{-1}: \mathbb{C}\cup\{\infty\} \rightarrow \mathbb{C}\cup\{\infty\}$ with $z\mapsto f^{-1}(z)=\frac{z-i}{z+i}$ (with $f\circ f^{-1}=\mathrm{Id}$).

\subsection{Hyperbolic Delaunay triangulations}

\begin{figure}
\centering
\includegraphics[bb=0 0 949 463, width=0.7\textwidth]{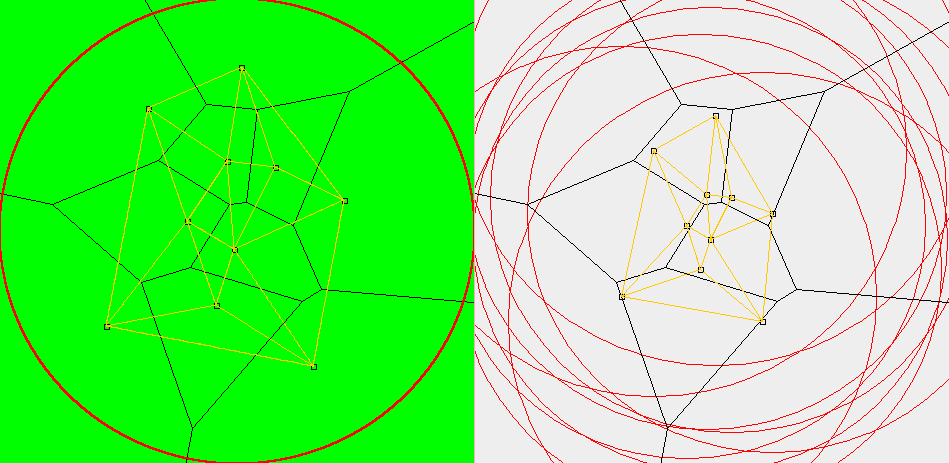}

\caption{
\label{fig:del}
Hyperbolic Delaunay and regular triangulations obtained as dual structures of Voronoi diagrams.
The regular triangulation (right) is embedded in the Klein model of hyperbolic geometry.}
\end{figure}

The  power diagram equivalent to the hyperbolic Voronoi diagram admits a dual regular triangulation~\cite{powerdiagrams-1987,compgeom-1998,bvd} as depicted in Figure~\ref{fig:del}.
This straight line triangulation can be embedded in Klein disk model of projective geometry.
Furthermore, the bisector of two points $\v{p}$ and $\v{q}$ is perpendicular to the geodesic linking $\v{p}$ to $\v{q}$.
In particular, the two circular arcs meet orthogonally in the Poincar\'e conformal representations.
In Klein disk model, they also meet perpendicularly but since the representation is not conformal, it makes a different Euclidean angle.
However, we can easily compute the angle between two line segments in Klein model by considering their chords.
Indeed, the ideal points located on the boundary circle are the same in the  Klein model and the Poincar\'e disk model. 

\subsection{Weighted and $k$-order hyperbolic Voronoi diagrams\label{sec:weight}}

Power diagrams are weighted (ordinary) Voronoi diagrams that can be obtained by projecting vertically the polytope defined by the intersection of half-spaces~\cite{powerdiagrams-1987} generated by lifting points to the paraboloid $z=\dotproduct{\v{x}}{\v{x}}$ of $\mathbb{R}^{d+1}$, taking the corresponding tangent plane and translating them vertically according to their weights.
A similar lifting/projection mechanism holds for hyperbolic Voronoi diagram in the upper-plane model (the proof relying on pencils of spheres~\cite{DMT92a} is omitted)
\begin{theorem}
The 2D hyperbolic Voronoi diagram on the upper plane can be obtained by first computing the intersection of 3D halfspaces implied by the paraboloid lifting, then projecting horizontally the edges of the 3D polytope to the paraboloid, and finally projecting vertically these curves on the plane. 
\end{theorem}

Furthermore, as shown in~\cite{bvd}, $k$-order and weighted affine diagrams are also affine diagrams that can therefore be computed equivalently as power diagrams. These power diagrams can be built themselves as the projection of a $(d+1)$-dimensional polytope obtained as the intersection of half-spaces tangent to the paraboloid and translated vertically by a corresponding weight.

\section{Applications of the hyperbolic Voronoi diagram\label{sec:app}}

\subsection{Browsing image collections in the hyperbolic disk\label{sec:ui}}

\begin{figure}
\centering
\includegraphics[bb=0 0 601 661, width=0.35\textwidth]{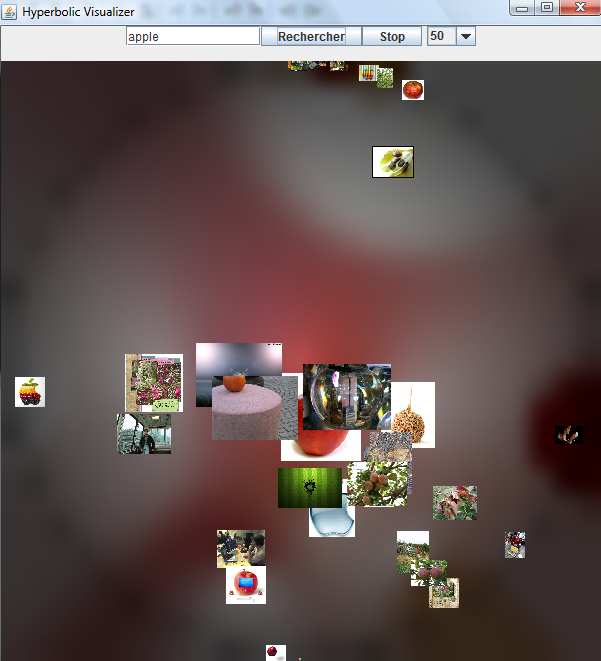}\ \ 
\includegraphics[bb=0 0 601 664, width=0.35\textwidth]{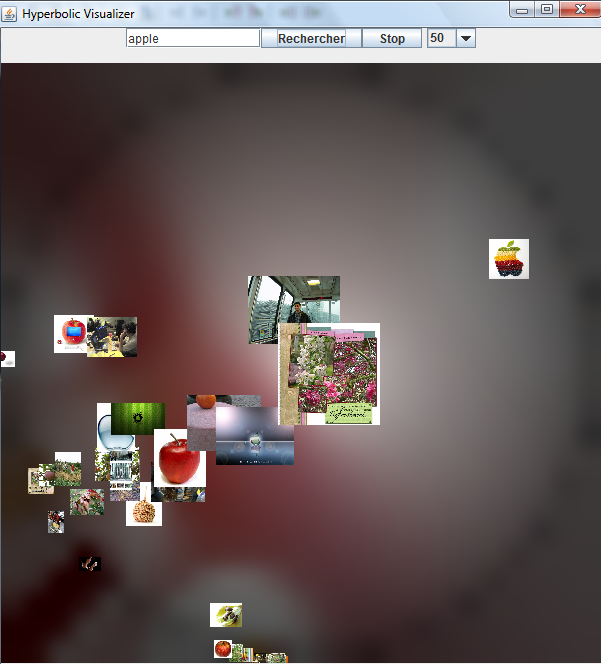}
\caption{
\label{fig:browser}
Snapshot of the image browser applet in the Poincar\'e hyperbolic disk.
This bounded representation of hyperbolic space greatly helps in navigating among large image collections. 
Users enter keywords from which internet images are automatically aggregated via the Yahoo! API. 
The left and right snapshots are related by a M\"obius transform driven by mouse user-interface interactions.}
\end{figure}

Figure~\ref{fig:browser} shows a snapshot of our interactive image search and browser application.
We use the fact the   Poincar\'e/Klein disk models allow one to display the full hyperbolic plane into a bounded disk fitting the display dimension.
The application asks for text keywords and query web search engines to retrieve large collections of images.
These images are analyzed using various feature extractions (e.g., histograms) and represented in the 2D hyperbolic space using Sammon clustering technique\footnote{See \url{http://www.techfak.uni-bielefeld.de/~walter/h2vis/}} that optimizes the non-linear embedding so as to preserve pairwise distances~\cite{hyperbolicimagebrowsing-2006} of high-dimensional extracted feature vectors.  

The Poincar\'e/Klein disk models of the hyperbolic space are not uniquely defined as it depends on the chosen origin and orientation around the disk center: The selected {\it viewpoint}~\cite{optimaltransformations-2001}, also called {\it focus}~\cite{focushvd-1995}. 
From the user standpoint, it is important to be able to interactively change the focus so as to better visually depict the structures of the image collection.
In particular, two useful user-interface operations involve computational geometry tasks

\begin{itemize}
\item Image selection: Given a mouse click, report the closest anchored image,
\item Group selection: Given several mouse clicks, report the smallest enclosing ball containing all anchored images falling within the mouse selection.
\end{itemize}

The first operation is a nearest neighbor query in the hyperbolic geometry.
The second operation is useful for re-centering the hyperbolic disk model using the viewpoint  so as to improve the display area of the selected images.
Since rectangular shapes does {\it not} visualize rectangularly in either Poincar\'e/Klein disk, we prefer to manipulate circles that are rasterized as circles in Poincar\'e conformal disk and ellipsis in the Klein non-conformal disk. 

The   transformations for re-centering the viewpoint of the disk models are M\"obius transformations.

{\bf Closest image and nearest neighbor queries}.
To answer the first specific UI operation of finding the nearest neighbor for a given mouse-click query, we build the Klein Voronoi diagram.
This yields a 2D straight line partition of the disk space (i.e., an arrangement) on which we perform point location in logarithmic time~\cite{compgeom-1998}.
We may recenter the viewpoint of the disk model at the selected image, if wished.

{\bf Group selection and smallest enclosing ball}. 
To address the second UI primitive, we first need to compute the smallest enclosing ball, and then recenter the viewpoint of the disk model so as to coincide with the circumcenter of that enclosing ball.
The smallest enclosing ball may be computed from the {\it furthest Voronoi diagram} that is an affine diagram in Klein disk in $O(n\log n)$ time, following~\cite{skyum}.
This is sub-optimal as we may rather use a simple linear randomized algorithm originally proposed by Welzl~\cite{Welzl91} for Euclidean geometry and recently extended to dually flat spaces~\cite{nn2008}.
Indeed the combinatorial optimization problem is of type LP (linear programming).
On the plane, we need to define the two basic primitives: 
The hyperbolic ball passing through two points, and the  hyperbolic ball passing through three points.
The first case is solved by computing the intersection point of the Klein bisector with the line segment joining these two points.
The second case considers the intersection any two of the three bisectors.
Once the circumcenter of the smallest enclosing ball is computed, we finally recenter the viewpoint of the disk model using a corresponding M\"obius transformation.

\begin{figure}
\centering
\includegraphics[bb=0 0 406 381, width=0.35\textwidth]{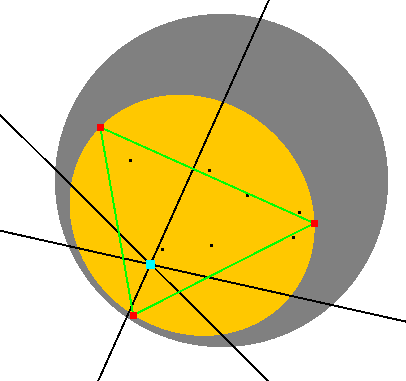}
\caption{\label{fig:hball} Smallest enclosing ball of a point set in the Klein non-conformal disk model: affine bisector and smallest enclosing ball deformed as an ellipsis.}
\end{figure}

\section{Conclusion}
We have shown that the  hyperbolic Voronoi diagram in the Klein projective disk model is merely an affine diagram that can be conveniently computed as an equivalent power diagram.
This crucial observation led us to a framework for performing other useful computational geometry tasks such as answering nearest neighbor queries or  computing smallest enclosing balls. We illustrated how these techniques come in handy for designing user interfaces of an interactive image browser application.

\vskip 0.2cm
\parindent 0cm {\bf Acknowledgments}\\
\parindent 0cm {\small
Part of this research has been financially supported by ANR-07-BLAN-0328-01 (GAIA) and
DIGITEO 2008-16D (GAS). 
We thank Cyprien Pindat for implementing in Java an earlier version of the image browser application in the Poincar\'e disk model.
}


\end{document}